\documentclass{aa}     

\usepackage{graphicx}
\usepackage{txfonts}

\newcommand{\Mpc}{$h^{-1}$\thinspace Mpc}

\def\apjl{ApJL} 
\def\apjs{ApJS}

\begin{document}

\title{Environmental Enhancement of DM Haloes}

\author {M. Einasto\inst{1}, I. Suhhonenko\inst{1,2},
P. Hein\"am\"aki\inst{1,2}, J. Einasto\inst{1}, E. Saar\inst{1}}

\authorrunning{M. Einasto et al.}

\offprints{M. Einasto }

\institute{ Tartu Observatory, EE-61602 T\~oravere, Estonia
\and
Tuorla Observatory, V\"ais\"al\"antie 20, Piikki\"o, Finland 
}

\date{ Received   2004 / Accepted ...  }

\titlerunning{EEH}


\abstract{ We study the properties of dark matter haloes of a LCDM
model in different environments.  Using the distance of the 5th
nearest neighbour as an environmental density indicator, we show that
haloes in a high density environment are more massive,
richer, have larger radii and larger velocity dispersions than haloes
in a low density environment. Haloes in high density regions move
with larger velocities, and are more spherical than haloes in low 
density regions. In addition, low mass haloes in the vicinity of the most
massive haloes are themselves more massive, larger, and have larger
rms velocities and larger 3D velocities than low mass haloes far from
massive haloes.  The velocities of low mass haloes near massive haloes
increase with the parent halo mass.  Our results are in agreement with
recent findings about environmental effects for groups and clusters of
galaxies from deep (SDSS and LCRS) surveys.


}

\maketitle

\section{Introduction}

Recent analyses of deep redshift surveys of galaxies, as the Las
Campanas Redshift Survey (LCRS), the Sloan Digital Sky Survey (SDSS) and
the two-degree-field (2dF) Galaxy Redshift Survey have demonstrated
the presence of environmental enhancement of groups and clusters of
galaxies -- groups and clusters of galaxies in high density regions are
richer, more massive, more luminous and have larger velocity
dispersions than groups and clusters of galaxies in low density
regions (Einasto et al. \cite{je2003a},
\cite{je2003b}, \cite{env}, \cite{lgs}, Ragone et al. \cite{r04}).

The goal of the present paper is to study the properties of dark
matter (DM) haloes
in LCDM simulations in various environments and to compare the
properties of these haloes with observational results for groups and
clusters of galaxies.

In the next Section we describe the numerical model and the halo
finding procedure used in our analysis.  Then we describe the method
to calculate environmental densities around haloes, and study the properties
of haloes in different environments.  In the last two Sections we give
a discussion and summary of our results.

The colour figures and the three-dimensional distribution of the
simulated haloes can be seen at the home page of Tartu Observatory
({\tt http://www.aai.ee/$\sim$maret/cosmoweb.html}), where 
we also present a catalogue of DM halos.

\section{The N-body model and the DM haloes}

For the present study we use a flat cosmological model with the
parameters obtained from the WMAP microwave background anisotropy
experiment (Bennett et al. \cite{bennet}): the total matter density
$\Omega_m=0.27$, the baryonic density $\Omega_b=0.044$, the vacuum
(dark) energy density (the cosmological constant)
$\Omega_{\Lambda}=0.73$, the Hubble constant $h=0.71$ (here and
throughout this paper $h$ is the present-day Hubble constant in units
of 100 km s$^{-1}$ Mpc$^{-1}$) and the rms mass density fluctuation
parameter $\sigma_8=0.84$.

The simulations were performed using the Multi Level Adaptive Particle
Mesh code (MLAPM, Knebe et al. \cite{knebe}).  This code uses adaptive
mesh techniques and adaptive force softening; new sub-grids are
created in regions where the density exceeds a specified
threshold. This scheme 
allows to improve considerably the spatial
resolution of the particle-mesh code without loss in the mass resolution,
similarly to the ART code (Kravtsov and Klypin \cite{kratso}).  It is
known that two-particle relaxation may enhance the fraction of
particles in clusters.  Adaptive force softening is found to prevent
mass segregation caused by two-body relaxation, while in other
algorithms (etc. GADGET), based on fixed force softening, two-body
relaxation is more prominent (Binney and Knebe \cite{binney}).

We use the results of a simulation run in a cube of 200~\Mpc\ size,
using a $256^3$ mesh and the same number of particles. Each particle
has a mass of $3.57\times 10^{10} h^{-1} {\rm M}_{\sun}$.  The
transfer function was computed using the COSMICS code by
E.~Bertschinger ({\tt http://arcturus.mit.edu/cosmics/}).  For
selection of dark matter haloes we used the FoF (Friends-of-Friends)
algorithm (Zeldovich, Einasto, and Shandarin \cite{zes82}).  

We shall use the linking length 0.23 in units of the mean particle
separation, which approximately corresponds to the matter density
contrast $\delta n/n$ = 80.  This value was used in compiling the
catalogue of the Las Campanas Loose Groups of Galaxies, hereafter LCLG
(Tucker et al. \cite{tuc}, Hein\"am\"aki et al. \cite{hei:hei}). It is
substantially lower than the spherical collapse model prediction for
virialized objects, $\delta n/n=178\Omega_m^{-0.6}$ (White, Efstathiou
and Frenk \cite{whi}).

Too large a linking length may lead to overestimation of the masses of
the simulated groups since particles outside of the virialized core
are included in the groups.  To avoid possible effects of unbound
groups in simulations, we used the virial condition
$E_r=E_{kin}/|E_{pot}|<0.5$ ($E_{pot}$ is the potential energy and
$E_{kin}$ the kinetic energy of a group) for groups to be included in
our final group catalogue. Moreover, to avoid too small groups we
choose only groups which included more than 100 dark matter particles,
imposing a minimum halo mass of $3.57\times 10^{12} h^{-1} {\rm
M}_{\sun}$.  The catalogue contained 5355 dark matter haloes, and
after applying the virial condition, 5219
dark matter haloes remained in our final catalogue.

\begin{figure}[ht]
\centering
\resizebox{0.45\textwidth}{!}{\includegraphics*{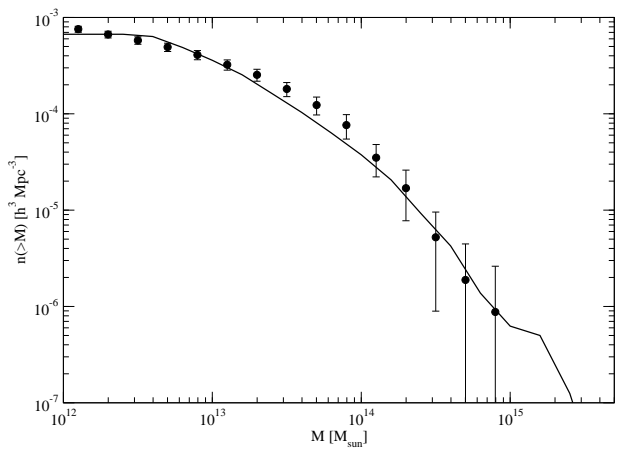}}
\hspace{2mm}
\caption{The observed mass function and the simulated mass function
for haloes; the filled circles show the LCLG result and the solid line
describes our simulations.
}
\label{fig:1}
\end{figure}

Fig.~\ref{fig:1} shows the cumulative mass function for dark matter haloes,
compared with observational data (LCLG). The mass function of the LCLG
(with Poisson error bars) is shown by filled circles. For a more
detailed description of selection effects in the LCLG sample and of
the determination of the LCLG mass function see Tucker et
al. (\cite{tuc}) and Hein\"am\"aki et al. (\cite{hei:hei}).  The solid
line shows the mass function of our simulations, with halo masses
determined by the sum of the particle masses in the halo.  We see that
the mass functions of the LCLGs and of the simulation are rather
similar. Less massive haloes in our simulation correspond to galaxy
groups; the most massive haloes correspond to rich clusters of
galaxies.

\section{Haloes in different environments}

\subsection{Environmental densities around haloes}
In the present work we
determine spatial densities around haloes, using the distance to the
halo's 5th neighbour halo. We shall use this distance as an
environmental parameter, 
When we use this distance as the
density indicator, we always see the characteristic scales of the
features we study.  This method has been used before to find
environmental (projected or spatial) densities around galaxies (Caon
and Einasto 1995, Goto et al.  \cite{goto03}), around clusters
(Schuecker et al.  \cite{schu01}, Einasto et al. \cite{e04a}) and
even around superclusters of galaxies (Einasto et al. \cite{e1997}).

Using the 5th neighbour distance we shall divide the haloes into four
environment regions as described in Table ~\ref{tab:t1}.  The
neighbour distance intervals are chosen to have approximately the same
number of haloes in each region.  The region H23.1 has the closest
neighbours and the highest environmental densities, the class H23.4 has the most
distant neighbours and the lowest environmental 
densities.

\begin{figure*}[ht]
\centering
\resizebox{0.45\textwidth}{!}{\includegraphics*{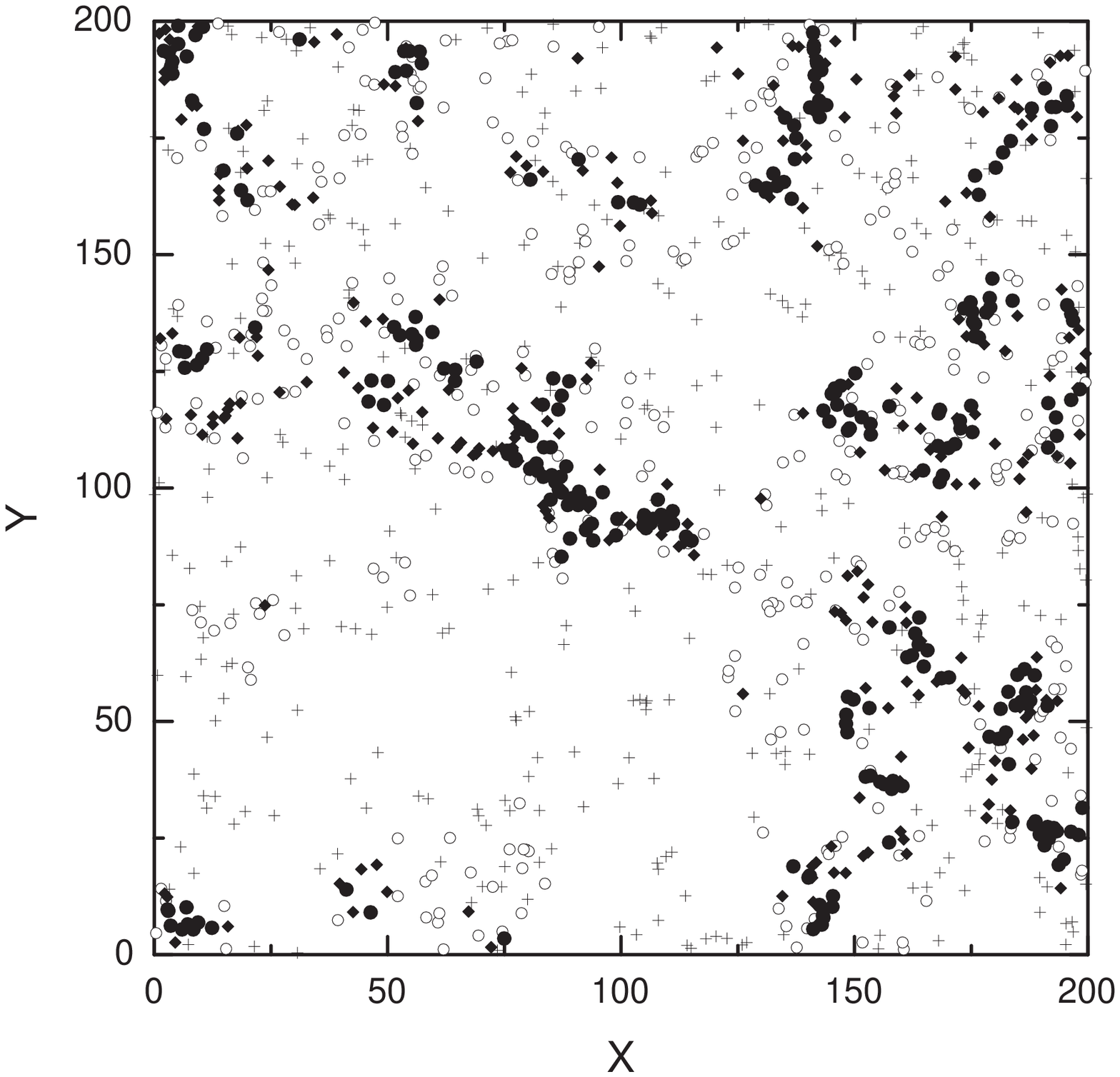}}
\hspace{2mm}
\resizebox{0.45\textwidth}{!}{\includegraphics*{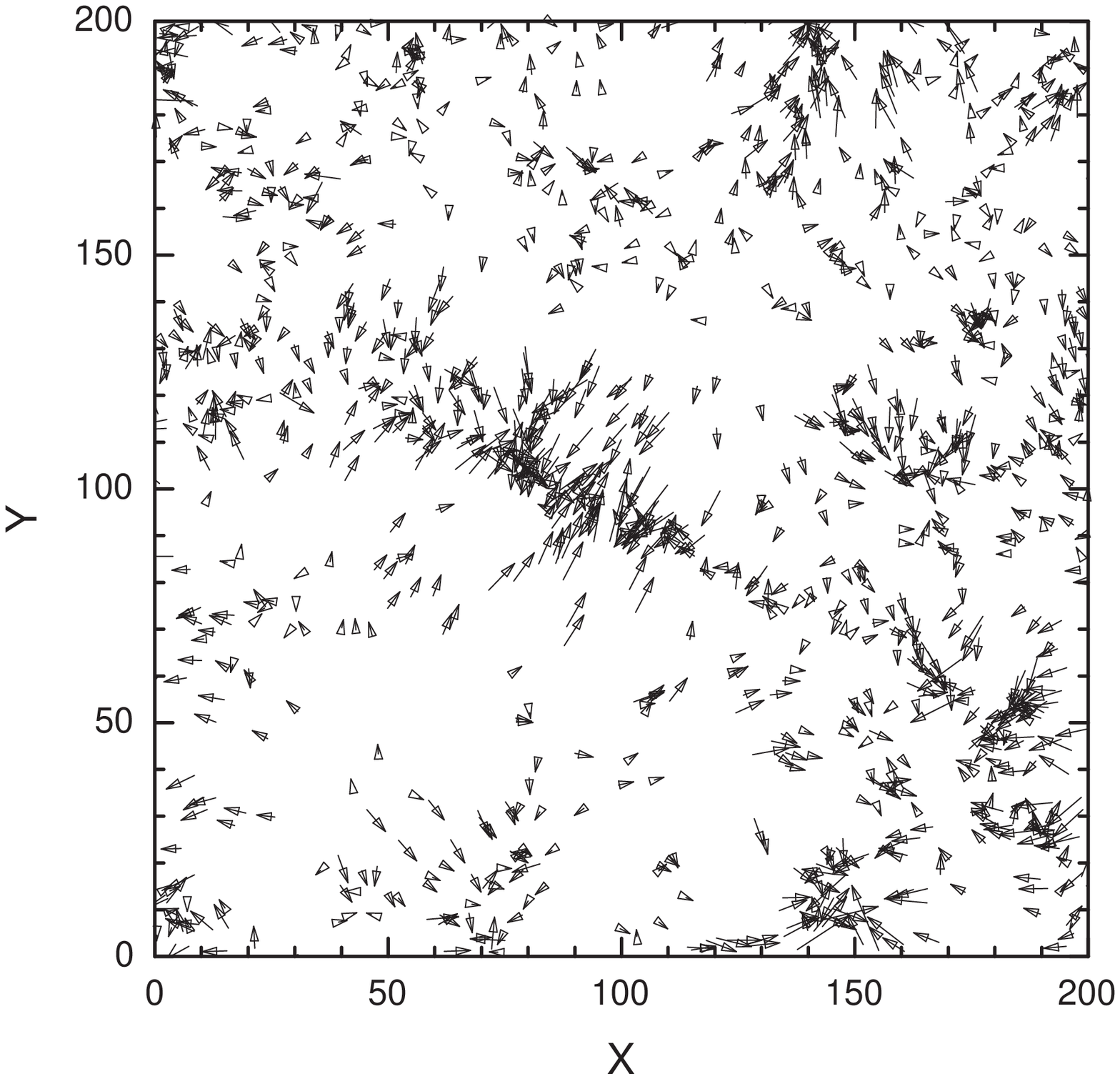}}
\hspace*{2mm}\\
\caption{The spatial distribution of haloes in regions of different
density in a slice of thickness of 50 \Mpc\ (in the Z-direction); 
 the Z interval is 150--200 \Mpc.  
The left panel shows the spatial distribution of
haloes, the right panel shows the velocity vectors of haloes. Filled
large circles denote haloes in the highest density environment (sample
H23.1 in Table ~\ref{tab:t1}), filled small diamonds stand for the
sample H23.2, open circles for the sample H23.3, and crosses denote
haloes in the lowest density environment, sample H23.4.
}
\label{fig:2}
\end{figure*}

\subsection{Two richness classes of haloes\label{sec:CN-DN}}

In Einasto et al. (\cite{env}) we showed that the properties of the LCLGs
in the neighbourhood of rich clusters are enhanced in comparison with
the properties of the LCLGs in average.  Moreover, in Einasto et
al. (\cite{lgs}) we showed that the LCLGs in superclusters are more
massive than loose groups of galaxies that do not belong to
superclusters.

To test this result using LCDM haloes, we divided our sample of LCDM
haloes into two richness classes: the most massive haloes that
correspond to rich clusters of galaxies, and less massive haloes that
correspond to loose groups of galaxies. Here we used the mass threshold $M
\geq 6.3\times 10^{13} h^{-1} {\rm M}_{\sun}$ (see Hein\"am\"aki et
al. \cite{hei:hei}).  Einasto et al. (\cite{env}) also used this threshold
to determine the population of Abell-class clusters among the LCLGs.

Once two subsamples with different masses were determined, we found
for the most massive haloes their 5 closest neighbours among less
massive haloes (with masses $M < 6.3\times 10^{13} h^{-1} {\rm
M}_{\sun}$).  Then we divided the population of low mass neighbours
into the populations of close neighbours (CN) and distant neighbours
(DN), using a threshold distance of 6 \Mpc. This distance was used in
Einasto et al. (\cite{env}) to define the populations of loose groups
around rich clusters of galaxies.

\section{The properties of haloes in different environment}
 
\subsection{Haloes in high and low density regions}

The spatial distribution of haloes in our $L=200$~\Mpc\ simulation box
is shown in Fig.~\ref{fig:2}; (a  50 \Mpc\ thick slice in the
Z-direction; color figures of all slices are
given in our web page together with animations).  This figure shows
haloes located in regions of different densities, as quantified by
the 5th nearest neighbour distance.  In general, in high density regions 
haloes are bigger and more massive than haloes located in low density
regions. The velocities of haloes in high density regions
are also larger than the velocities of haloes in low density regions. 
The velocity vectors of haloes are directed toward rich
filaments and massive haloes -- mass flows from voids into filaments,
which act as attractors.  In extreme void regions, where the haloes
are most isolated, the haloes almost do not move; they also have low
masses.

\begin{table*}
\begin{center}
\tabcolsep 5pt

\caption{Median and upper quartile (in parentheses) values of halo properties
in high and low density environments.}

\begin{tabular}{lcccccccccc}
\hline 
Sample & $N_{\rm halo}$ & $D_1(N5)$& $D_2(N5)$ & $N_{\rm p}$ & 
$\log M_{\rm vir}$  & $R_{\rm v}$ & $\sigma$  & $V$  & $E$
& $ N_M $\\
\hline 
(1)& (2) & (3) & (4) & (5)& (6) & (7) & (8)& (9) & (10) & (11) \\

\hline 
H23.1 & 1315  &  0.0 &  7.9 & 409 (992)& 12.70 (13.17)& 0.40 (0.51)&
240 (380) & 480 (640) & 0.50 (0.63)& 101 \\ 
H23.2 & 1287  &  7.9 &  9.7 & 357 (816)& 12.59 (13.04)& 0.38 (0.50)&
210 (340) & 445 (585) & 0.51 (0.65)&  66 \\ 
H23.3 & 1340  &  9.7 & 12.2 & 309 (651)& 12.49 (12.92)& 0.37 (0.46)&
190 (300) & 400 (535) & 0.50 (0.64)&  37 \\ 
H23.4 & 1280  & 12.2 &      & 247 (463)& 12.32 (12.70)& 0.36 (0.44)&
160 (250) & 345 (470) & 0.52 (0.66)&  18 \\ 

\hline
\end{tabular}
\label{tab:t1}
\end{center}

The columns in the Table are as follows:

\noindent column 1: the halo sample number,

\noindent column 2: the number of haloes in the sample,

\noindent column 3: the smallest 5th nearest neighbour distance,

\noindent column 4: the largest 5th nearest neighbour distance,

\noindent column 5: the number of particles in a halo,

\noindent column 6: the virial mass of a halo (log),
$M_{vir}=\sigma^2R_{vir}/G$,

\noindent column 7: the virial radius of a halo 
(for equal mass particles), $R_{vir}=N^2/(\sum_{pairs}\frac{1}{r_{ij}})$,

\noindent column 8: the rms velocity of halo particles,
$\sigma=\sqrt{\sum_{i} v_i^2/N}$, where the squared velocity deviation
$v_i^2=(v_{ix}-v_{cx})^2+(v_{iy}-v_{cy})^2+(v_{iz}-v_{cz})^2$ 
is relative to the mean velocity of the halo and 
$v_{cx}, v_{cy}, v_{cz}$ are the components of the mean velocity.

\noindent column 9: the velocity of a halo,
$v_{halo}=\sqrt{v_{cx}^2+v_{cy}^2+v_{cz}^2}$

\noindent column 10: the eccentricity of a halo, $\varepsilon = 1 -
cc/aa $, where $cc$ is the semi-minor axis and $aa$ is the semi-major axis.

\noindent column 11:  the number of haloes with masses 
$M \geq  6.3\times 10^{13} h^{-1} {\rm M}_{\sun}$.

\end{table*}

\begin{figure}[ht]
\centering
\resizebox{0.45\textwidth}{!}{\includegraphics*{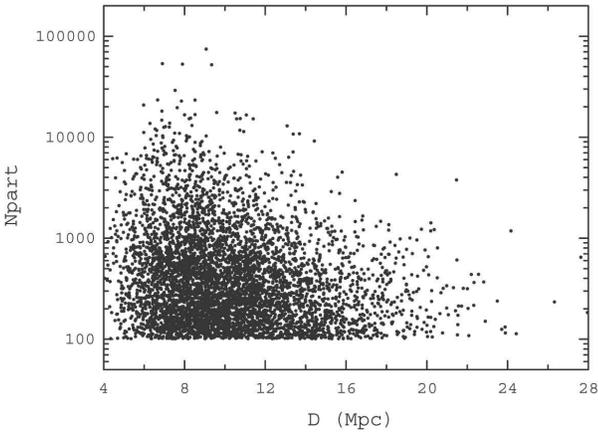}}
\hspace{2mm}
\caption{The number of particles in haloes versus the distance of the
5th nearest neighbour of the halo.}
\label{fig:3}
\end{figure}

Let us now analyse the properties of haloes in various environments in
more detail. We made a series of scatter plots, in which we plotted
the number of particles in haloes, halo's virial radii, masses and
velocity dispersions against the distance of the 5th nearest neighbour
of the halo, and calculated cumulative distributions of these
parameters (the number of particles in haloes, halo's virial radii, masses
and velocity dispersions) for haloes from different environments, and
the median and upper quartile values of the properties of haloes from
high and low density regions.  We have also estimated the statistical
significance of our results, using the Kolmogorov-Smirnov test.

We compare haloes from different environments in Table~\ref{tab:t1}.
The columns show median and upper quartile (in parentheses) values of halo
properties in high and low density environments.  In order to not to
overcrowd the paper with figures, we show only one scatter plot, for
the halo richness (Fig.~\ref{fig:3}).

\begin{figure*}[ht]
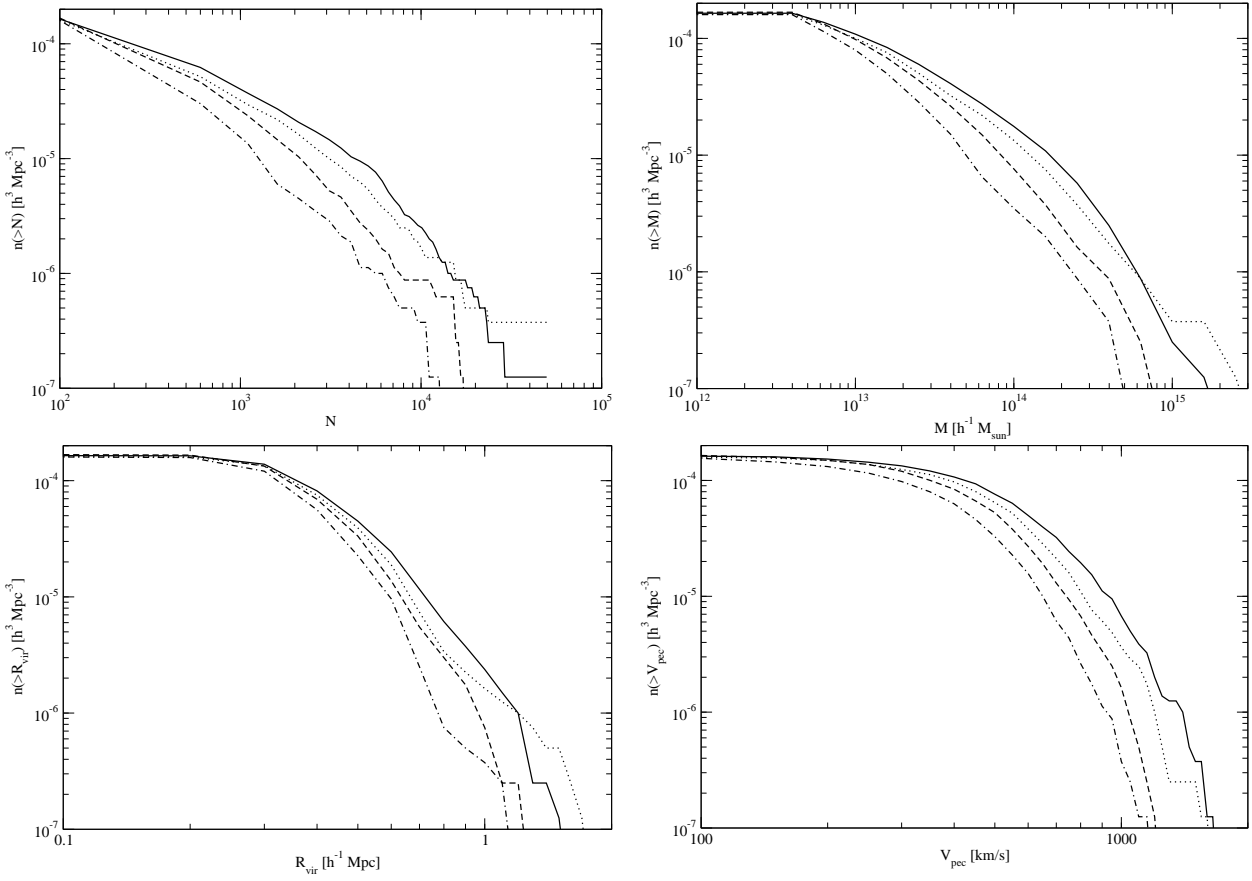

\centering
\resizebox{0.45\textwidth}{!}{\includegraphics*{Einasto4a.eps}}
\hspace{2mm} 
\raisebox{-5pt}{\resizebox{0.45\textwidth}{!}{\includegraphics*{Einasto4b.eps}}}
\hspace*{2mm}\\
\raisebox{-2pt}{\resizebox{0.45\textwidth}{!}{\includegraphics*{Einasto4c.eps}}}
\hspace{2mm} 
\resizebox{0.45\textwidth}{!}{\includegraphics*{Einasto4d.eps}}
\hspace*{2mm}\\
\caption{The cumulative distributions of various parameters of haloes
in different environments.  Upper left panel: cumulative richness.
Upper right panel: the mass functions.  Lower left panel: the
cumulative virial radius.  Lower right panel: the cumulative velocity.
Solid line -- the supercluster region H23.1, dotted and dashed lines --
the filament regions H23.2 and H23.3, and dot-dashed line -- the void region
H23.4.
}
\label{fig:4}
\end{figure*}

Fig.~\ref{fig:3} and Table~\ref{tab:t1} show that the properties
of haloes in high density regions are environmentally enhanced:
they are richer, larger,
more massive, and have larger velocity dispersions than haloes in low
density regions. The haloes in high density regions have larger
velocities than haloes in a low density environment.  In the lowest
density regions (low density filaments and voids) haloes almost do not
move. This is seen also in Fig.~\ref{fig:2}.

The cumulative mass function (MF) is defined as the number density of
clusters/groups above a given mass $M$, \mbox{$n(>M)$}.  We used the
same definition also for other properties of the haloes. Thus we get
the cumulative richness distribution, the cumulative virial radius
distribution, the cumulative rms velocity distribution, and the
cumulative peculiar velocity distribution.  We plot these functions
for haloes in different environments in Fig.~\ref{fig:4}.
These figures and Table~\ref{tab:t1}
show that haloes in a high density environment are
richer, they have larger masses, larger radii,
larger rms velocities and and larger
peculiar velocities than haloes in a low density environment.  For
example, at the density level $10^{-6}h^3\mbox{Mpc}^{-3}$ the H23.1
clusters are about 3 times richer than the H23.4 clusters.
The cluster class haloes (with mass $M \geq
6.3\times 10^{13} h^{-1} {\rm M}_{\sun}$) are preferentially located
in high density regions.  A few cluster class haloes, however, are
located in lower density  environments. This is due do the
selection procedure of the haloes -- the most massive haloes do not
have very close neighbours (they are halo's members), thus the local
density is suppressed in the neighbourhood of the halo.  This is seen
also in colour figures in our web page.

\begin{figure*}[ht]
\centering
\resizebox{0.45\textwidth}{!}{\includegraphics*{Einasto5a.eps}}
\hspace{2mm}
\raisebox{-15pt}{\resizebox{0.45\textwidth}{!}{\includegraphics*{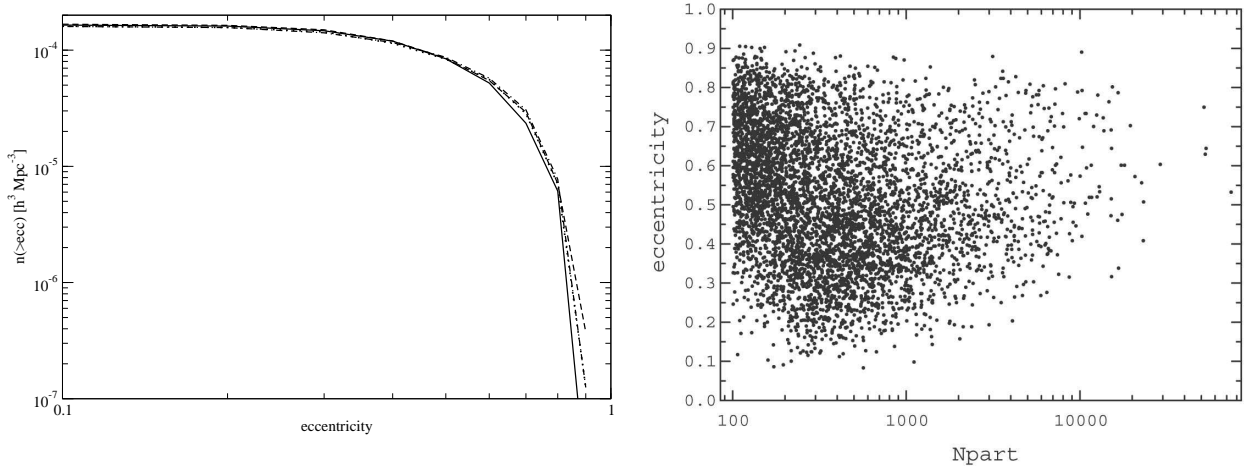}}}
\hspace*{2mm}\\
\caption{Left panel: the  cumulative distributions of eccentricities for
simulated haloes in different environments.
Solid line --- H23.1, dotted line --- H23.2, dashed line --- H23.3, 
and dot - dashed line --- H23.4. 
Right panel: The eccentricity of haloes versus the richness of haloes,
for all environments.
}
\label{fig:5}
\end{figure*}

We have determined also the eccentricities of DM-haloes.   
Fig.~\ref{fig:5} and Table~\ref{tab:t1} hint at the trend that
the eccentricities of haloes in high density regions seem to be smaller
than the eccentricities of haloes in low density regions. Haloes in high
density regions are more spherical than haloes in low density regions,
an evidence of a higher degree of virilization.

Fig.~\ref{fig:5} shows also, that the differences between the
eccentricities of haloes from different environments are smaller than
the differences between other parameters (distributions
on the Fig.~\ref{fig:5}, left panel, for samples
H23.2 and H23.3 even practically overlap).  The reason for that can be
seen in the right panel of Fig.~\ref{fig:5}: haloes of medium
richness (about $200 - 1000$ particles, mass $\leq 3.57\times 10^{13}
M_{\sun}$) have smaller eccentricities.  This is probably due to
several factors: haloes of medium richness are more evolved than very
poor haloes, this makes them more spherical. The shape of the richest
haloes is more elongated due to the influence of the large scale
structure around them (filaments). The environmental dependence of the
shapes of haloes is affected by these factors and is thus weaker than
other environmental dependences.

The Kolmogorov-Smirnov test shows that the differences between the
parameters of haloes from  high and  low density regions are
statistically significant at the 99\% significance level for all
parameters except eccentricities.  The differences between
eccentricities of haloes in the highest and the lowest density regions
are statistically significant at the 75\%
significance level according to the Kolmogorov-Smirnov test.

\subsection{Properties of haloes in the neighbourhood
of the most massive haloes}

\begin{figure*}[ht]
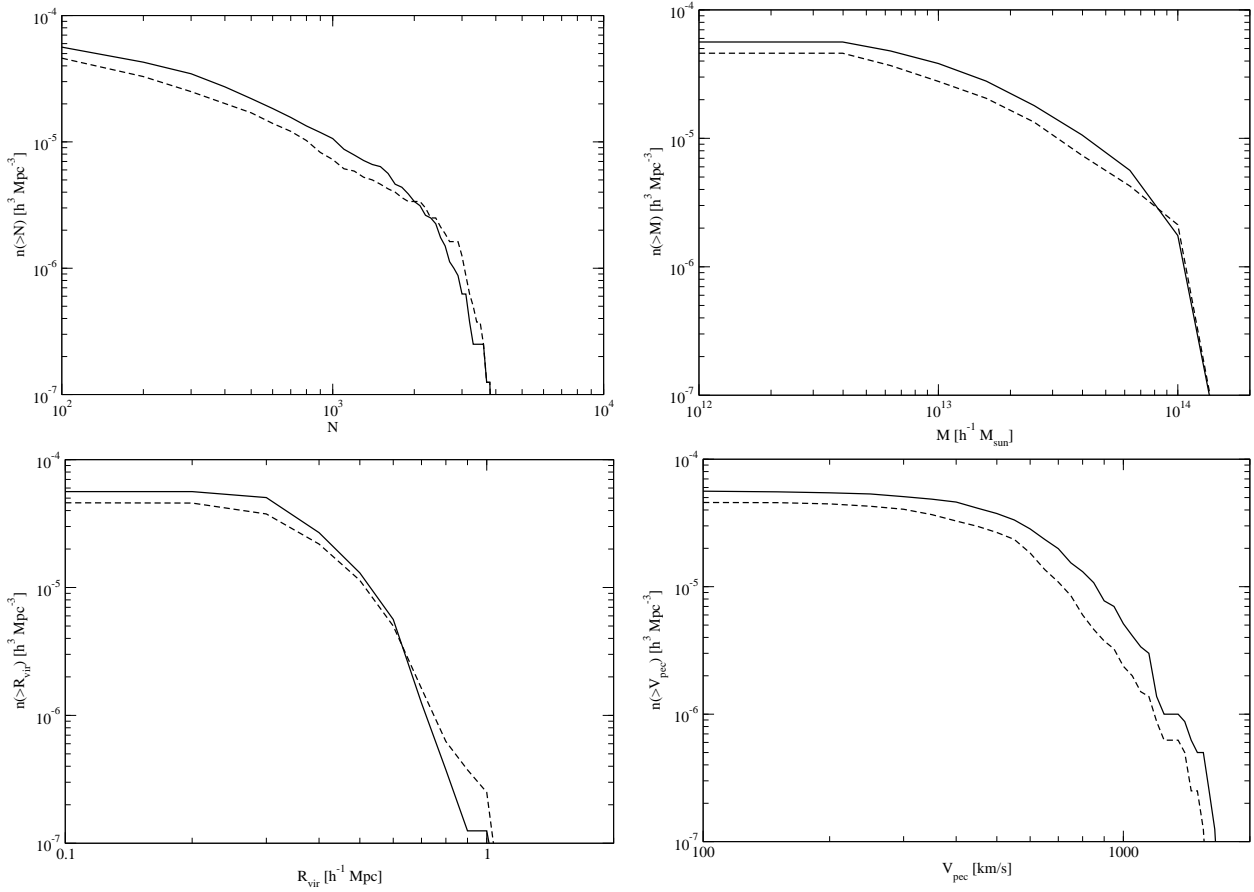

\centering
\resizebox{0.45\textwidth}{!}{\includegraphics*{Einasto6a.eps}}
\hspace{2mm} 
\raisebox{-5pt}{\resizebox{0.45\textwidth}{!}{\includegraphics*{Einasto6b.eps}}}
\hspace*{2mm}\\
\raisebox{-2pt}{\resizebox{0.45\textwidth}{!}{\includegraphics*{Einasto6c.eps}}}
\hspace{2mm} 
\resizebox{0.45\textwidth}{!}{\includegraphics*{Einasto6d.eps}}
\hspace*{2mm}\\
\caption{The cumulative distributions of various parameters for low
mass haloes in the vicinity of the most massive haloes.  Upper left
panel: cumulative richness.  Upper right panel: the mass functions.
Lower left panel: the cumulative rms velocity.  Lower right panel: the
cumulative peculiar velocity.  Solid line --- close neighbours, the sample CN,
dashed line --- distant neighbours, the sample DN.
}
\label{fig:6}
\end{figure*}

In Table~\ref{tab:t2} we give the median and upper quartile values of
the parameters of the halo populations CN and DN (close neighbours and
distant neighbours, see Sect.~\ref{sec:CN-DN}). In
Fig.~\ref{fig:6} we show  the cumulative
richness distributions, the mass functions,
the cumulative virial radius distributions,
and the cumulative peculiar velocity distributions for low mass haloes
around high mass haloes.
In Table~\ref{tab:t2} we also give the statistical significances of
the differences between the samples, according to the
Kolmogorov-Smirnov test.

Fig.~\ref{fig:6} and Table~\ref{tab:t2} show that the low mass haloes in
the close neighbourhood of high mass haloes are themselves richer,
more massive, have larger rms velocities and move faster than the low
mass haloes far from massive haloes.
The eccentricities of haloes in the neighbourhood of massive haloes
are smaller than the eccentricities of haloes farther away from
massive haloes.

We also checked whether the properties of low mass haloes in the
vicinity of massive haloes depend on the masses of high mass haloes.
Our calculations show that while the intrinsic properties of
low mass haloes do not seem to depend on the mass of the parent
massive halo, peculiar velocities of low mass haloes in the vicinity
of massive halo increase with the parent halo mass
(Fig.~\ref{fig:7}, left panel) -- 
an indication of the gravitational influence of massive haloes.

We obtain similar results if we use summed particle masses instead of
virial masses for the massive haloes.

Our calculations show that if we find all low mass neighbours around
high mass haloes within a 6 \Mpc\ sphere, (not just 5 closest
neighbours), then we obtain almost the same population (with about 470
member haloes) of low mass neighbours. Thus we used the five closest
neighbours, in accordance with our definition of environmental
densities around haloes.

\begin{table*}
\begin{center}
\tabcolsep 5pt

\caption{Median and upper quartile (in parentheses) values of the properties
of low mass haloes in the neighbourhood of massive haloes
.}

\begin{tabular}{lccccccccc}
\hline 
Sample & $N_{\rm halo}$ & $D_1(N5)$& $D_2(N5)$ & $N_{\rm p}$ & 
$\log M_{\rm vir}$  & $R_{\rm v}$ & $\sigma$  & $V$ & $E$
\\
\hline 
(1)& (2) & (3) & (4) & (5)& (6) & (7) & (8)& (9) & (10)  \\

\hline 
CN & 450  &  0.0 &  6.0 & 395 (774)& 12.73 (13.07)& 0.40 (0.50)& 240
(340) & 600 (770)& 0.51 (0.62) \\ 
DN & 368  &  6.0 &      & 339 (711)& 12.61 (13.00)& 0.40 (0.50)& 220
(320) & 550 (690)& 0.50 (0.65) \\ 
KS &      &      &      & 80\%     & 98\%         & 85\%       & 97\%
& 99\%     & 70\%        \\ 
\hline
\end{tabular}
\label{tab:t2}
\end{center}

The columns are as in Table~\ref{tab:t1}.  The last line shows the
significance levels of the difference of the distributions, by the
Kolmogorov-Smirnov test.
\end{table*}

\section{Discussion}

\subsection{Comparison with observations}

Already early studies of galaxies in different environments showed the
presence of strong morphology-density and luminosity-density relations
(Dressler \cite{dr80}, Postman \& Geller \cite{pg84}, Einasto \&
Einasto \cite{ee87}, Einasto \cite{e91a} and \cite{e91b}, Mo et
al. \cite{mexd92}).  Presently, these relations are studied
intensively, using new deep surveys of galaxies (Norberg et
al. \cite{n01}, \cite{n02}, Croton et al. \cite{cr04}, Kauffmann et
al. \cite{kau04}, Balogh et al. \cite{balogh04}, and Blanton et
al. ~\cite{blan04} among others).

However, the study of the properties of galaxy groups and clusters in
various environments is only in the beginning.  Observations show that
groups and clusters of galaxies in high density regions, in the
vicinity of rich clusters of galaxies and in superclusters, are more
massive and luminous and have larger velocity dispersions than loose
groups in average (Einasto et al. \cite{{je2003a}}, \cite{je2003b}, 
\cite{env}, \cite{lgs}, \cite{e04a}, 2004b).  This enhancement
extends to scales up to about 15--20~\Mpc\ around rich clusters.

Ragone et al. (\cite{r04}) determined groups of galaxies for the 2dF 
Galaxy Redshift Survey and studied the properties of these groups in the 
vicinity of rich clusters. This study confirms the results 
obtained in Einasto et al. (\cite{env}): groups in the vicinity of 
rich clusters of galaxies are themselves also richer and more massive
than groups in average.

\begin{figure*}[ht]
\centering
\resizebox{0.45\textwidth}{!}{\includegraphics*{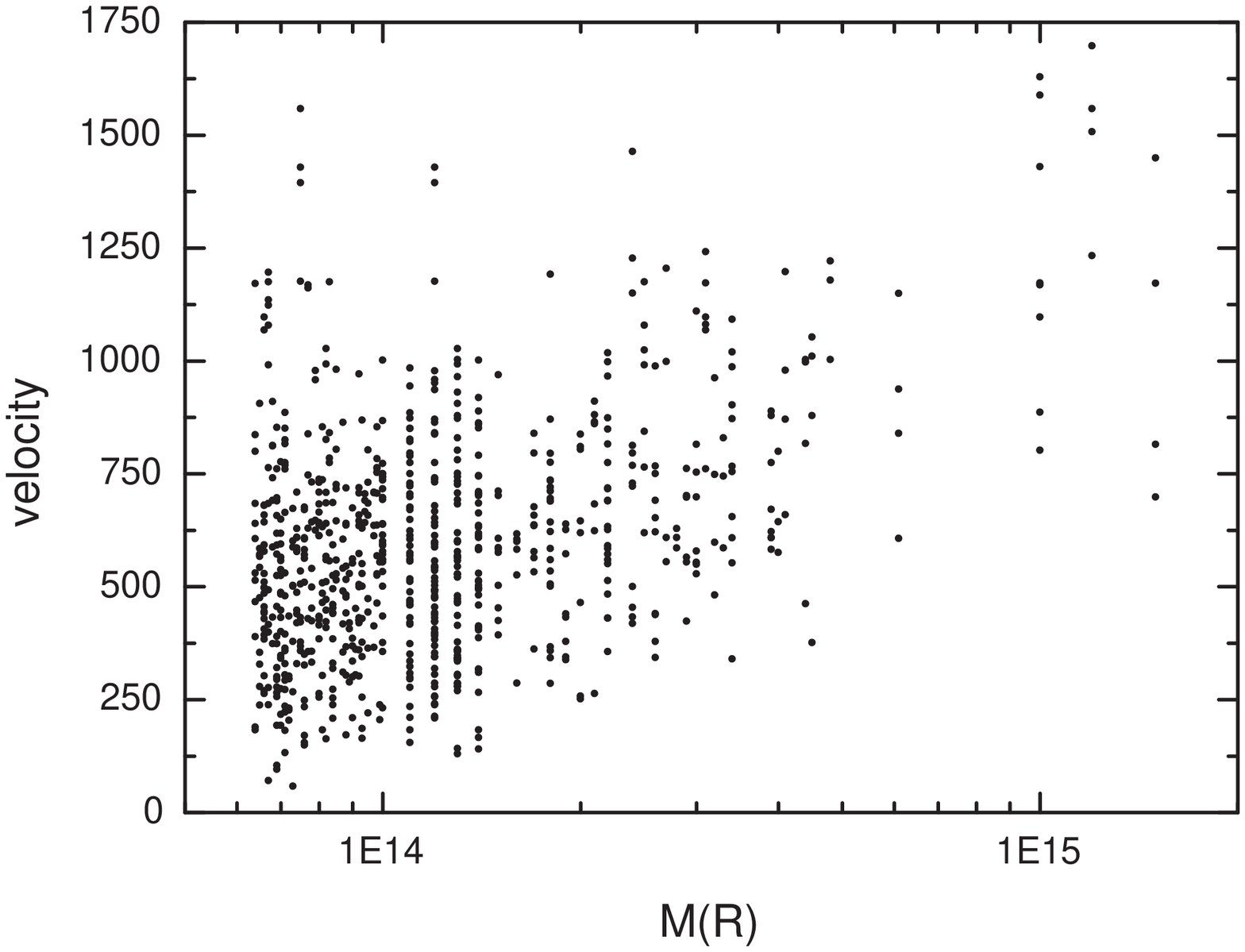}}
\hspace{2mm}
\resizebox{0.445\textwidth}{!}{\includegraphics*{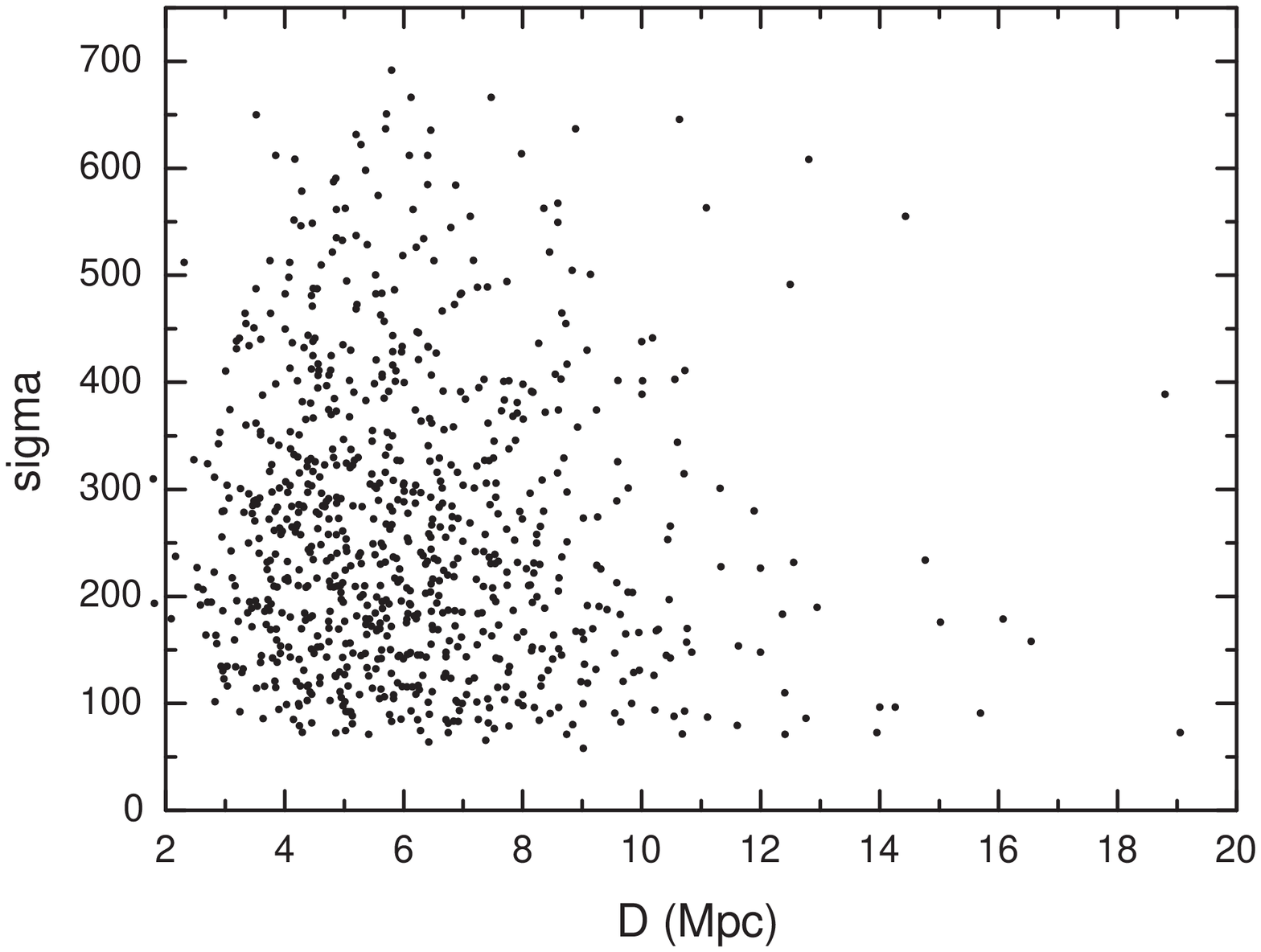}}
\hspace*{2mm}\\
\caption{Left: peculiar velocities (in km/s)nof low mass haloes 
around high mass haloes versus the mass of the high mass halo.
Right: the rms velocities (in km/s) of low mass haloes around high mass
haloes versus the distance of the low mass halo from the high mass
halo.
}
\label{fig:7}
\end{figure*}

In Einasto et al. (\cite{lgs}) we studied the environmental
enhancement of the LCLGs in superclusters. In particular, we calculated
for the LCLGs in superclusters the distance to the nearest Abell cluster
in supercluster and studied the properties of the LCLGs as a function of
the distance to the nearest rich cluster in supercluster. In analogy
to that test we plot the distribution of rms velocities of low mass
haloes in the vicinity of the massive haloes (Fig.~\ref{fig:7}, 
right panel).
Note the similarity of Fig.~\ref{fig:7}, right panel, 
and Fig.~5 in Einasto et
al. (\cite{lgs}): groups close to rich clusters have larger rms
velocities than groups far from rich clusters in a supercluster.

Several studies of the correlation function of nearby groups of
galaxies show that groups of higher mass are more strongly clustered
than groups on average (Giuricin et al. \cite{giu01}, Girardi et
al. \cite{gir00}, and Merchan et al. \cite{mer00}). Stronger
clustering is an indication that these groups could be located in high
density regions (Einasto et al. \cite{e1997}).

Plionis, Basilakos and Tovmassian (\cite{pli04}) showed that the
shapes of poor groups of galaxies depend on their richness, poorer
groups being more elongated than richer groups and clusters of
galaxies.  Observations show that very rich clusters of galaxies also
are elongated (see references in Plionis, Basilakos and Tovmassian
\cite{pli04}). This is in accordance with our results, showing that the
poorest haloes have larger eccentricities than haloes of medium
richness, and the most rich haloes again have larger eccentricities
than medium rich haloes.

Several studies of clusters of galaxies have provided evidence that
properties of rich clusters depend on their large scale environment
(Novikov et al. \cite{nov99}, Chambers et al. \cite{cham01}, Einasto
et al. \cite{e2001}, Schuecker et al. \cite{schu01} and Plionis and
Basilakos \cite{plb02}) up to a distance of about $20$~\Mpc. This
distance is close to the so-called ``pancake scale'' (Melott \&
Shandarin \cite{mel93}), and corresponds to the mean thickness of
superclusters (Einasto et al.  \cite{e1994}, \cite{e1997}, and
Jaaniste et al. \cite{ja98}).

Our results about the properties of haloes in different environments
are in accordance with these observational results.  The LCDM model
haloes in high density regions are richer, more massive etc than
haloes in low density environment. In particular, low mass haloes
around the most massive haloes are themselves also more massive and
have larger velocities than haloes farther away from massive haloes.

Our study shows that haloes move toward  massive
systems. In the vicinity of massive systems
the velocities of haloes are larger than in low density regions (voids).  
This is in accordance with
observational studies of velocity fields of galaxy clusters which show
that clusters are moving toward superclusters of galaxies (Hudson et
al. \cite{hud04} and references therein).

\subsection{Comparison with simulations}

Ragone et al. (\cite{r04}) determined groups of galaxies from
simulations by the Virgo Consortium, and studied the properties of
these groups in the vicinity of rich clusters. This study confirmed
the results obtained using observational data: groups in the vicinity
of rich clusters of galaxies are themselves also richer and more
massive than groups in average.

Ragone et al. (\cite{r04}) also found that the enhancement of the
properties of low mass haloes in the vicinity of massive haloes is a
strong function of the mass of these massive haloes, while we did not
detect such a strong dependence (except for 3D velocities). This
difference is probably due to the fact that in this test, Ragone et
al.  (\cite{r04}) used different host samples with increasing mass
limit for host haloes, while we did not change the host sample.

Suhhonenko (\cite{suhh}) has demonstrated using different N-body
simulations that in simulated superclusters more massive clusters are
located in the central regions of superclusters.

Gottl\"ober et al. (\cite{got:got}) and Faltenbacher et
al. (\cite{fal:fal}) analysed high-resolution simulations of formation
of galaxies, groups, and clusters, and found a significant enhancement
of the mass of haloes in the environment of other haloes. This effect
is especially significant at scales below 10 $h^{-1}$Mpc. Therefore,
environmental enhancement of the halo mass is a direct evidence for
the process of the hierarchical formation of galaxy and cluster
haloes. Halos inside clusters formed earlier than haloes in low
density filaments and are more evolved (Gottl\"ober, Klypin and
Kravtsov \cite{got00}, Gottl\"ober et al. \cite{got03}).

Kasun and Evrard (\cite{kas04}) found that the shapes of LCDM haloes
do not depend on the large scale (supercluster) environment of
haloes. However, the small hint of the such correlation seen in our
study may ensue from the different definitions of the environmental
density.

The finding that the richest haloes have larger eccentricities is in
accordance with the recent results by Hopkins, Bahcall and Bode
(\cite{hbb04}) who also found that clusters are aligned with one
another, in accordance with several observational studies (see
references in Hopkins, Bahcall and Bode \cite{hbb04}) and the results by
Colberg et al. (\cite{col99}) who showed that the formation of
a cluster in simulations is governed by its surrounding large scale
structure, and the properties of clusters are linked to the properties
of structures around them. More massive clusters have more filaments
than less massive clusters (Colberg, Krughoff and Connolly
\cite{col04a}). This finding is in agreement with our result about the
environmental dependencies of halo's properties. Moreover,
observations show that rich superclusters usually have a
multi-branch structure and these superclusters contain more
luminous clusters than poor superclusters (Einasto et
al. \cite{je2003b}).

Bahcall, Gramann and Cen (\cite{bgc94}) and Colberg et
al. (\cite{col00}) found evidence that rich clusters have larger
peculiar velocities than poor clusters, and clusters in high density regions
have larger peculiar velocities than clusters in low density regions.

In supercluster-void network haloes move toward density enhancements,
where the velocities of haloes are larger than velocities of haloes in
a low density environment: rich systems become richer and the fraction
of matter in voids decreases (see also Einasto et
al. \cite{ev94}). Halos in the lowest density regions have the
smallest masses in agreement with the recent results about the mass
function of haloes in voids (Colberg et al. \cite{col04b}).

Our present paper extends the earlier findings about the environmental
enhancement of haloes and shows that haloes in higher density
environment, in general, are richer, more massive and have larger
velocity dispersions than haloes in low density environment.

A possible explanation of the environmental enhancement of galaxy
systems in high-density regions was investigated by Einasto et
al. (\cite{e04a} and \cite{e04b}).  These studies show that dynamical
evolution in high-density regions is determined by the high overall
mean density that speeds up the clustering of particles.  Therefore,
in high density regions clustering starts early and continues
until the present.  The haloes that populate high density regions are
themselves also richer, more massive and have larger velocities than
the haloes in low density regions.  In low density filaments that cross
voids, as well as in the outer low density regions of high density
systems, the mean density decreases and thus the evolution is slow, and
in these regions haloes themselves are also poor, less massive and have
small velocities.

\section{Conclusions}

We studied the properties of dark matter haloes in different environments.
Our main results are as follows.

\begin{itemize}
\item{} We composed a catalogue of haloes found using the MLAPM code and
the FoF algorithm with a linking length of 0.23 in the
units of mean particle separation. This catalogue is available 
from our web pages.

\item{} Our analysis shows that haloes in a high density environment
are richer, more massive, have larger virial radii,
have larger rms velocities, have larger peculiar velocities and
are more spherical than haloes in a low density environment.

\item{} Low mass haloes in the vicinity of high mass haloes are themselves
richer, more massive, slightly more spherical and have larger rms
velocity dispersions than low mass haloes farther away from high mass
haloes. The larger the mass of parent halo, the larger are the
velocities of low mass haloes in the vicinity of high mass haloes.

\item{} Our study indicates the importance of the role of a high density
environment, which affects the properties (formation and evolution) of
galaxy systems.  In high density regions haloes formed earlier, and
are more evolved than haloes in low density regions, in accordance
with the scenario of the hierarchical formation of the structure in
the Universe.

\end{itemize}

\begin{acknowledgements}

We thank Mirt Gramann for stimulating discussions. The present study was
supported by the Estonian Science Foundation  grant 4695 and  by 
the Estonian Research and Development  Council   grant  TO 0060058S98. 
P.H. was supported by the Jenny and Antti Wihuri 
foundation.

\end{acknowledgements}

\end{document}